\documentclass[epj]{svjour}
\usepackage{graphicx}

\usepackage{amssymb}
\usepackage{amsmath}

\usepackage[colorlinks]{hyperref}

\AtBeginDocument{%
    \hypersetup{
      urlcolor     = blue, 
      linkcolor    = black, 
      citecolor   = blue, 
      pdfborder={0 0 0},
      urlbordercolor={1 1 1},
      citebordercolor={1 1 1}
    }
    }

\begin{document}
\title{The quantum optical description of three experiments involving non-linear optics using a graphical method}
\author{Stefan~Ataman
\thanks{\emph{Alternative e-mail address:} stefan.ataman@gmail.com}%
}                     
%
%
\institute{\email{ataman@ece.fr}}
\date{Received: date / Revised version: date}
%
\abstract{In this paper we describe and thoroughly discuss three
reported experiments in quantum optics (QO) involving
interferometers and non-linear crystals. We show that by using a
graphical method and an over-simplified model of the parametric
down-conversion process, we arrive to explain all the important
results reported in the respective papers. Indistinguishability is
discussed in the case of separable/non-separable (i.e. entangled)
quantum systems and our interpretation is sometimes at variance
with the one given by the authors reporting the experiments.
%
\PACS{
      {PACS-key}{describing text of that key}   \and
      {PACS-key}{describing text of that key}
     } 
} 

\titlerunning{The quantum optical description of three experiments using a graphical method}
\authorrunning{S. Ataman}

\maketitle
%

\section{Introduction}

Indistinguishability in Quantum Mechanics plays a key role in many
experiments. Indeed, adding quantum amplitudes associated with the
indistinguishable paths rather than probabilities, allows one to
explain the observed interference phenomena
\cite{HOM87,Pit96,Ou90F,Wal02}.

The fact that in a Mach-Zehnder interferometer single quanta of
light are in a coherent superposition of being simultaneously in
both arms has many counter-intuitive consequences, among others
the ``click by click'' buildup of the interference pattern
\cite{Gra86} or the the so-called ``interaction-free
measurements'' \cite{Eli93,Kwi95}.

The non-linear process of spontaneous parametric down-conversion
(SPDC) \cite{Bur70} provides pairs of highly entangled photons
\cite{Kly67,Hon85,Rub94-Gri97-Kel97,Kwi95b}. It had been used
hitherto in many experiments in quantum optics
\cite{HOM87,Pit96,Ou90F,Shi88-Rar90,Kwi93,Ou90a,Ou90b,Zou91,Wang91,Lem14}.
One could emphasize the fundamental experiments demonstrating the
non-locality of quantum mechanics \cite{Shi88-Rar90,Kwi93}.


In an experiment by Ou, Wang, Zou and Mandel (proposed in
\cite{Ou90a} and reported in \cite{Ou90b}), two non-linear
crystals fed by the same pump laser through a beam splitter
perform down-conversions. By slightly varying the position of the
beam splitter (and therefore the relative phase of the pump field
between the crystals), an interference pattern is recovered. The
authors model this by considering a coherent superposition between
the output wavepacket and the vacuum state. We will arrive at the
same results using a simpler model, not needing to resort to this
``phase memory'' of the vacuum.


Another experiment using two  non-linear crystals showing a quite
counter-intuitive behavior was reported by Zou, Wang and Mandel in
\cite{Zou91} and its theoretical model further refined in
\cite{Wang91}. This time, although the idler beams of the two
non-linear crystals are discarded, they seem to play a major role
in the interference of the other beams. The apparent paradox of
this experiment will be analyzed later on.


Based on the principle of the previous experiment, the rather
surprising idea of imaging an object with undetected photons was
recently reported in an experiment \cite{Lem14}, where a series of
small objects has been imaged using undetected (discarded) photons
and arrays of photon counters that detected photons not crossing
the object's path.


In this paper we describe three experiments involving non-linear
crystals using the graphical method introduced in \cite{Ata14b}.
Although a very simple model for the non-linear process of SPDC is
used, all important phenomena reported in the respective papers
can be accounted for. The experiments are thoroughly discussed,
showing were an explanation differing from the authors' one can be
given.


This paper is organized as follows. In Section
\ref{sec:field_op_transf_theoretical} we give a theoretical
justification for the field operator transformations and
supplement the graphical method introduced in \cite{Ata14b} with
the inclusion of the non-linear quantum optical phenomenon of
spontaneous parametric down-conversion. The experiment of Ou,
Wang, Zou and Mandel \cite{Ou90b} is described using the graphical
method in Section \ref{sec:ou_wang_zou_mandel_1990} and the
results compared to the ones experimentally obtained. The more
counter-intuitive experiment of Zou, Wang and Mandel \cite{Zou91}
is discussed in Section \ref{sec:zou_wang_mandel_experiment}.
Using a similar experimental setup, the imaging of an object with
undetected photons \cite{Lem14} is described and discussed in
Section \ref{sec:zeilinger_experiment}. Finally, conclusions are
drawn in Section \ref{sec:conclusions}.


\section{Field operator transformations in quantum optics}
\label{sec:field_op_transf_theoretical}

Throughout this paper, we shall consider an input state derived
from the pump laser. Since we select only events where a
down-conversion takes place, we can write the input state as
\begin{equation}
\label{eq:psi_input_a_p_plus_void}
\vert\psi_{in}\rangle=\vert1_p\rangle=\hat{a}_p^\dagger\vert0\rangle
\end{equation}
where $\vert1_p\rangle$ denotes a Fock state with one photon in
mode (port) $p$, $\hat{a}_p^\dagger$ denotes the input field
creation operator and $\vert0\rangle$ is the vacuum state. In
order to find the output state, we need an operator transformation
function $g$ connecting the input creation field operator to the
output ones,
\begin{equation}
\label{eq:operator_p_function_of_output_ops}
\hat{a}_p^\dagger=g\left(\hat{a}_s^\dagger,\hat{a}_{s'}^\dagger,\hat{a}_{i}^\dagger,\hat{a}_{i'}^\dagger\right)
\end{equation}
where $\hat{a}_k^\dagger$ is the creation operator for the output
port $k$ with, $k=\{s,{s'},i,{i'}\}$. Therefore, at least
formally, the output state can be written as
\begin{equation}
\label{eq:psi_out_f_vacuum_monochromatic} \vert\psi_{out}\rangle
=g\left(\hat{a}_s^\dagger,\hat{a}_{s'}^\dagger,\hat{a}_{i}^\dagger,\hat{a}_{i'}^\dagger\right)\vert0\rangle
\end{equation}
In \cite{Ata14b} a graphical method allowing the fast computation
of field operator transformation has been introduced for linear
optical systems. We extend this method to non-linear
transformations by including the SPDC process.

The SPDC process produces two entangled single photon wave packets
\cite{Kly67,Rub94-Gri97-Kel97} that are generally very
short\cite{HOM87} (hence non-monochromatic). The input beam,
generally called the pump ($p$) is split into two output beams,
called the signal ($s$) and the idler ($i$). Energy
($\hbar\omega$) and momentum ($\hbar\boldsymbol{k}$) conservation
relations impose $\omega_p=\omega_s+\omega_i$ and
$\boldsymbol{k}_p\approx\boldsymbol{k}_s+\boldsymbol{k}_i$ (the
latter being also called ``phase matching'' condition). In this
paper, a very simplified model of the SPDC process will be
employed, where this process converts a monochromatic input photon
from the pump field into two (equally monochromatic) photons in
the signal, and, respectively, idler outputs.  From the
perspective of QO, the field operator transformations in the SPDC
process are
\begin{equation}
\label{eq:SPDC_field_transf}
\hat{a}_p^\dagger=\gamma\hat{a}_s^\dagger\hat{a}_i^\dagger
\end{equation}
where $\gamma$ is a parameter connected to the $\chi^{(2)}$
non-linearity of the optical medium. 
For simplicity, throughout this paper we will consider\footnote{In
practice we have $\gamma\sim10^{-6}$. However, the choice of
$\gamma=1$ in our analysis can be justified by the fact that we
only consider events where a down-conversion actually takes
place.} $\gamma=1$, hence equation \eqref{eq:SPDC_field_transf}
transforms the input state \eqref{eq:psi_input_a_p_plus_void} into
\begin{equation}
\vert\psi_{out}\rangle=\hat{a}_s^\dagger\hat{a}_i^\dagger\vert0\rangle
=\vert1_s1_i\rangle
\end{equation}
at the output of the non-linear crystal.

\begin{figure}
\centering
\includegraphics[width=1in]{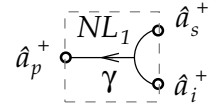}
\caption{The graphical description of the non-linear process of
spontaneous parametric down-conversion. The input (creation) field
operator $\hat{a}_p^\dagger$ can be written as a product of the
two output (creation) field operators, $\hat{a}_s^\dagger$ and
$\hat{a}_i^\dagger$.} \label{fig:graphical_method_SPDC}
\end{figure}

\section{The experiment of Ou, Wang, Zou and Mandel}
\label{sec:ou_wang_zou_mandel_1990}

\begin{figure}
\centering
\includegraphics[width=3in]{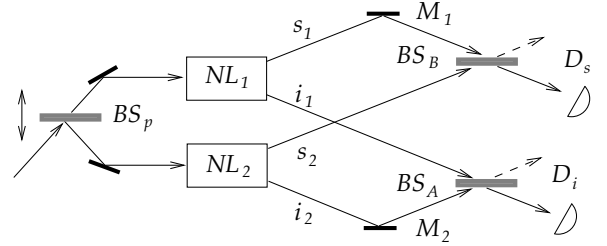}
\caption{The Ou, Wang, Zou and Mandel experiment \cite{Ou90b}. The
interfering paths $s_1-s_2$ and $i_1-i_2$ are made as equal as
possible. The beam splitter $\text{BS}_p$ is slightly moved as
indicated by the arrow causing a phase difference between the two
pump fields.} \label{fig:ou_wang_zou_mandel_experiment}
\end{figure}

The experiment performed by Ou, Wang, Zou and Mandel \cite{Ou90b}
is depicted in Fig.~\ref{fig:ou_wang_zou_mandel_experiment}. A
pump laser beam is divided by the beam splitter $\text{BS}_p$
between two non-linear crystals, denoted $NL_1$ and $NL_2$. The
signal and, respectively, idler beams from the two crystals are
brought together at beam splitters $\text{BS}_B$ and,
respectively, $\text{BS}_A$. The optical path lengths from the
crystals to the beam splitters were made as equal as possible. The
mixed signal ($s_1$ and $s_2$) and, respectively, idler ($i_1$ and
$i_2$) photons are then sent to the photo-detectors $D_s$, and,
respectively, $D_i$. The two-photon coincident detection rates
were measured in respect with the displacement of $\text{BS}_p$
(Fig.~3 from \cite{Ou90b}) and a sinusoidal variation was found.

\begin{figure}
\centering
\includegraphics[width=3in]{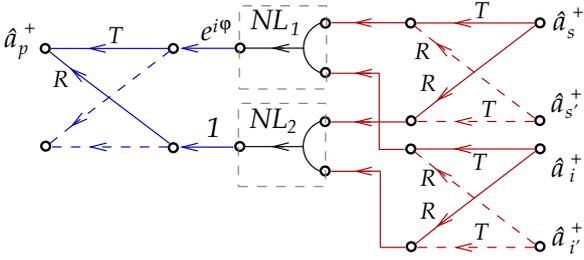}
\caption{(Color online) The graphical description of the Ou, Wang,
Zou and Mandel experiment. The beam splitters are represented as
``butterflies''. Ports connected to dashed paths are either unused
or will be traced out in the density matrix operator. The
$\text{e}^{i\varphi}$ coefficient models the path length
difference between $\text{BS}_p-NL_1$ and $\text{BS}_p-NL_2$.}
\label{fig:graphical_method_ou_wang_zou_mandel_color}
\end{figure}

In Fig.~\ref{fig:graphical_method_ou_wang_zou_mandel_color} the
same experiment in described using the graphical method
\cite{Ata14b}. The three beam splitters $\text{BS}_p$,
$\text{BS}_A$ and $\text{BS}_B$ (assumed identical), are depicted
by the three ``butterflies''. The beam splitters are assumed to
have a transmission (reflection) coefficient $T$ ($R$). Energy
conservation imposes the well-known constraints $\vert
T\vert^2+\vert R\vert^2=1$ and $T^*R+TR^*=0$ \cite{Lou03}. The
variable path length between $\text{BS}_p$ and the two non-linear
crystals is modelled by the phase shift\footnote{We wish to
express the input creation operator in respect with the output
field operators, therefore all arrows point ``backwards in time''.
This is why the phase shift is $\text{e}^{i\varphi}$ and not
$\text{e}^{-i\varphi}$.} $\text{e}^{i\varphi}$. The non-linear
crystals are depicted using the graph introduced in
Fig.~\ref{fig:graphical_method_SPDC}.

In order to express $\hat{a}_p^\dagger$ in respect with the output
field operators we examine the paths connecting them in
Fig.~\ref{fig:graphical_method_ou_wang_zou_mandel_color}. From the
crystal $NL_1$  to $\hat{a}_p^\dagger$ we have a factor
$T\text{e}^{i\varphi}$. The signal (upper) port of $NL_1$ is
connected to $\hat{a}_{s}^\dagger$ with a coefficient $T$ and to
$\hat{a}_{s'}^\dagger$ with a coefficient $R$. The idler (lower)
port of $NL_1$ is connected to $\hat{a}_{i}^\dagger$ via a
coefficient $T$ and to $\hat{a}_{i'}^\dagger$ via a coefficient
$R$. Putting together these contributions, one ends up with
$T\text{e}^{i\varphi}\times\left(T\hat{a}_s^\dagger+R\hat{a}_{s'}^\dagger\right)
\times\left(T\hat{a}_i^\dagger+R\hat{a}_{i'}^\dagger\right)$.
Similar arguments allow us to add the contribution of the second
crystal and we obtain the field operator transformation
\begin{eqnarray}
\hat{a}_p^\dagger=T\text{e}^{i\varphi}\left(T\hat{a}_s^\dagger
+R\hat{a}_{s'}^\dagger\right)\left(T\hat{a}_i^\dagger+R\hat{a}_{i'}^\dagger\right)
\nonumber\\
+R\left(R\hat{a}_s^\dagger+T\hat{a}_{s'}^\dagger\right)\left(R\hat{a}_i^\dagger+T\hat{a}_{i'}^\dagger\right)
\end{eqnarray}
or, in other words, we got the function $g$ from equation
\eqref{eq:operator_p_function_of_output_ops}.
All we have to do now is apply this result to
equation~\eqref{eq:psi_out_f_vacuum_monochromatic}. After some
simplifications and assuming all beam splitters to be balanced
($T=1/\sqrt{2}$ and $R=i/\sqrt{2}$ \cite{Lou03}), we get
\begin{eqnarray}
\label{eq:psi_out_ou_wang_zou_mandel_entangled}
\vert\psi_{out}\rangle=\frac{1}{\sqrt{2}}\Big(\sin\left(\varphi'\right)\vert1_s1_i\rangle
+\cos\left(\varphi'\right)\vert1_{s'}1_i\rangle
 \nonumber\\
+\cos\left(\varphi'\right)\vert1_{s}1_{i'}\rangle
-\sin\left(\varphi'\right)\vert1_{s'}1_{i'}\rangle\Big)
\end{eqnarray}
where we ignored a common phase factor and denoted
$\varphi'=\varphi/2-\pi/4$.

We could now  simply project the output state vector given by
equation \eqref{eq:psi_out_ou_wang_zou_mandel_entangled} into the
$\vert1_s1_i\rangle$ state and obtain the probability of
coincident counts $P_{s-i}$. However, for the sake of generality,
we shall employ the density matrix approach, outlined in Appendix
\ref{sec:appendix_rho_operator}. Therefore, we define the output
density matrix operator
$\hat{\rho}_{out}=\vert\psi_{out}\rangle\langle\psi_{out}\vert$
and trace it over the two unused outputs ($s'$ and $i'$) ending up
with the partial trace operator
$\hat{\rho}_{s,i}=\text{Tr}_{s',i'}\left\{\hat{\rho}_{out}\right\}$
given in equation \eqref{eq:partial_rho_s_i_Ou_et_al}. The
probability of coincident counts\footnote{Throughout this paper we
assume ideal photo-detectors.} at the detectors $D_s$ and $D_i$ is
now easily computed as
\begin{eqnarray}
P_{s-i}=\text{Tr}\left\{\hat{a}_{s}^\dagger\hat{a}_{s}\hat{a}_{i}^\dagger\hat{a}_{i}\hat{\rho}_{s,i}\right\}
=\frac{1}{4}\left(1+\sin\left(\varphi\right)\right)
\end{eqnarray}
and we find indeed an interference pattern while varying the
position of the beam splitter $\text{BS}_p$. If we replace the
phase $\varphi$ by $k_pz_p$ where $k_p$ is the wavenumber of the
pump laser and $z_p$ is the path length difference, we end up with
the interference pattern $P_{s-i}\sim1+\cos\left(k_pz_p\right)$
i.e. the interference fringes are periodic with the pump field
frequency. This is consistent with what is reported in the paper
(Fig.~3 in \cite{Ou90b}).

The authors conclude ``[$\ldots$] we have demonstrated that the
light produced in the down-conversion process carries information
about the pump phase through entanglement with the vacuum.''
\cite{Ou90b}. While this is one way to explain the experiment, it
might also be discussed using the simple model used throughout
this section. Fock states have indeed ill-defined phases, but we
can give an operational meaning to a phase difference between the
quantum superposition of two Fock states.

It is interesting to note that for the probability of single
counts at, say, detector $D_s$ one obtains
\begin{eqnarray}
\label{eq:Ps_singles_rate_no_interf}
P_{s}=\text{Tr}\left\{\hat{a}_{s}^\dagger\hat{a}_{s}\hat{\rho}_{s}\right\}
=\frac{\sin^2\left(\varphi'\right)+\cos^2\left(\varphi'\right)}{2}=\frac{1}{2}
\end{eqnarray}
where $\hat{\rho}_{s}=\text{Tr}_i\left\{\hat{\rho}_{s,i}\right\}$
and no interference fringes can be found on varying $\varphi$.
This is about to change in the next section, where a modified
version of this experiment will be discussed.

\section{The experiment of Zou, Wang and Mandel}
\label{sec:zou_wang_mandel_experiment}

In \cite{Zou91}, Zou, Wang and Mandel performed an experiment that
has been dubbed ``mind boggling'' \cite{Gre93}. The
counter-intuitive part in this experiment comes from the fact that
some photons interfere or not conditioned on the
distinguishability of other, \emph{undetected} photons.

\begin{figure}
\centering
\includegraphics[width=3in]{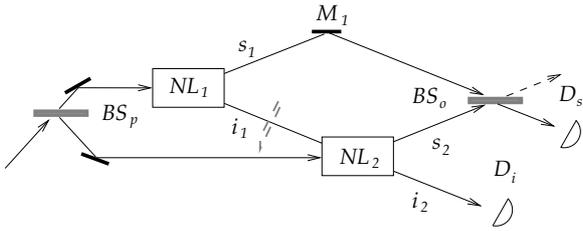}
\caption{The Zou, Wang and Mandel experiment \cite{Zou91}. The
idler beam from the non-linear crystal $NL_1$ is aligned so that
it will overlap with the idler beam from $NL_2$. Thus,
indistinguishability of the idler beams is assured at the detector
$D_i$.} \label{fig:zou_wang_mandel_experiment}
\end{figure}

The experiment is depicted in
Fig.~\ref{fig:zou_wang_mandel_experiment}. Similar to the previous
experiment, a pump laser is divided by a beam splitter between two
non-linear crystals. The two signal beams ($s_1$ and $s_2$) are
brought together in the beam splitter $\text{BS}_o$ and a
photo-detector $D_s$ is placed at its output. The idler beam of
the first crystal ($i_1$) is passed through the second one (the
crystals are transparent, therefore attenuation is negligible) and
superposed on the idler beam $i_2$. A detector $D_i$ is placed at
the output idler beam. Given the configuration of this experiment,
any detection at $D_i$ is unable to pinpoint the origin of the
light quantum. What Zou, Wang and Mandel observed was an
interference pattern on monitoring the singles detection rate at
$D_s$, while varying the path length difference by moving the beam
splitter $\text{BS}_o$. Up to this point, no ``mind boggling''
features showed up in this experiment: the singles rate is the
result of an interference between the undistinguishable paths of
an interferometer composed of $\text{BS}_p$, $NL_1$, $NL_2$ and
$\text{BS}_o$.

However, if we block the beam $i_1$ (or simply make it
distinguishable in respect with $i_2$), the interference pattern
disappears. This result is rather surprising since the idler beams
do not participate to the interference happening at the beam
splitter $\text{BS}_o$.

\begin{figure}
\centering
\includegraphics[width=2.8in]{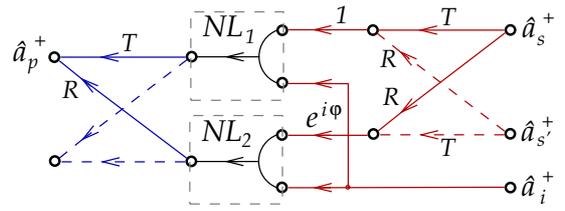}
\caption{(Color online) The graphical description of the Zou, Wang
and Mandel experiment \cite{Zou91}. Ports connected to dashed
paths are either unused or will be traced out in the density
matrix operator. The two idler beams have been connected together
and they end up at the port $\hat{a}_i^\dagger$.}
\label{fig:graphical_method_zou_wang_mandel_indist_color}
\end{figure}

In Fig.~\ref{fig:graphical_method_zou_wang_mandel_indist_color} we
describe this experiment using the graphical method. As before,
the two beam splitters (assumed identical) are depicted by the two
``butterflies'' while the path length difference in the
interferometer formed by the two non linear crystals and the beam
splitter $\text{BS}_o$ is modelled by the phase shift
$\text{e}^{i\varphi}$. By a simple inspection of the graph we can
express contribution to the the field operation transformation of
the upper path (the one containing $NL_1$) as
$T\times\left(T\hat{a}_s^\dagger+R\hat{a}_{s'}^\dagger\right)\times\hat{a}_i^\dagger$.
Adding the contribution of the second crystal yields the final
result
\begin{equation}
\hat{a}_p^\dagger=\left(T^2+\text{e}^{i\varphi}R^2\right)\hat{a}_s^\dagger\hat{a}_i^\dagger
+TR\left(1+\text{e}^{i\varphi}\right)\hat{a}_{s'}^\dagger\hat{a}_i^\dagger
\end{equation}
and for the case of balanced beam splitters ($T=1/\sqrt{2}$ and
$R=i/\sqrt{2}$) we end up with the output state
\begin{equation}
\label{eq:psi_out_zou_wang_mandel_non-entangled}
\vert\psi_{out}\rangle=-\sin\left(\varphi/2\right)\vert1_s1_i\rangle
+\cos\left(\varphi/2\right)\vert1_{s'}1_i\rangle
\end{equation}
where we ignored a common phase factor. We compute again the
density matrix
$\hat{\rho}_{out}=\vert\psi_{out}\rangle\langle\psi_{out}\vert$
and partially trace it over the unused output ($s'$) yielding
\begin{eqnarray}
\hat{\rho}_{s,i}=\sin^2\left(\varphi/2\right)\vert1_s1_i\rangle\langle1_s1_i\vert
+\cos^2\left(\varphi/2\right)\vert0_{s}1_{i}\rangle\langle0_{s}1_{i}\vert
\qquad
\end{eqnarray}
The probability of coincident counts at the detectors $D_s$ and
$D_i$ is immediately obtained as
\begin{eqnarray}
P_{s-i}=\text{Tr}\left\{\hat{a}_{s}^\dagger\hat{a}_{s}\hat{a}_{i}^\dagger\hat{a}_{i}\hat{\rho}_{s,i}\right\}
=\sin^2\left(\varphi/2\right)
\end{eqnarray}
and not surprisingly one finds an interference pattern. If we
write the phase shift $\varphi$ as $k_sz_s$, where $k_s$ is the
wavenumber in the $s$ mode and $z_s$ denotes the path length
difference of the signal beams to the beam splitter, the
coincident rate becomes
$P_{s-i}\sim1/2(1-\cos\left(k_sz_s\right))$, consistent with the
observed frequency of the interference fringes (Fig.~2 in
\cite{Wang91}).

However, the counter-intuitive feature of this experiment does not
come from the coincident counts. Indeed, by computing the singles
detection rate at detector $D_s$ one gets
\begin{eqnarray}
\label{eq:P_s_singles_interference}
P_{s}=\text{Tr}\left\{\hat{a}_{s}^\dagger\hat{a}_{s}\hat{\rho}_{s}\right\}
=\sin^2\left(\varphi/2\right)
\end{eqnarray}
where $\hat{\rho}_{s}=\text{Tr}_i\left\{\hat{\rho}_{s,i}\right\}$
and this time, contrary to equation
\eqref{eq:Ps_singles_rate_no_interf}, there is an interference
pattern present on varying $\varphi$.

A natural question arises: why in this experiment the singles
detection rate at $D_s$ yielded an interference pattern while in
equation \eqref{eq:Ps_singles_rate_no_interf} it did not? We could
argue that in both cases, a detection at $D_s$ or $D_i$ resulted
from two indistinguishable paths, therefore if we apply the dictum
``ignorance is interference'' this result should be rather
surprising.

A logical guess would be that the field from $i_1$ induced
down-conversions in the crystal $NL_2$. As the authors pointed out
\cite{Zou91}, this is tenable if the field intensities are high
enough. However, the same phenomenon is present for very low
intensities \cite{Wis00}. It is found that the down-conversions in
$NL_1$ and $NL_2$ are spontaneous (thus uncorrelated) even when an
idler photon from $i_1$ crosses $NL_2$. This conclusion is
strengthened in \cite{Lem14}, where the authors experimentally
showed that induced emission in $NL_2$ plays no role.

The authors conclude in their paper that ``in quantum mechanics
interference is always a manifestation of the intrinsic
indistinguishability of the photon paths, in which case the
corresponding probability amplitudes add. [$\ldots$] once the
$i_1$, $i_2$ connection is broken it becomes feasible, in
principle, to determine from the counts [$\ldots$] $D_i$ whether
the detected signal photon comes from $NL_1$ or $NL_2$, and this
destroys the interference'' \cite{Zou91}. While this affirmation
is certainly not wrong, it cannot contain the whole story; if it
would, equation \eqref{eq:Ps_singles_rate_no_interf} should have
shown an interference pattern, too.

It is well known that a composite system whose global wavevector
can be written in a factorized form allows the separate analysis
of each subsystem. This is not the case for an entangled system,
where each subsystem cannot be discussed separately. Therefore,
besides indistinguishability, one has to consider if each
subsystem (the signal and idler beams in this case) can be
analyzed separately\footnote{A more involved discussion about
non-separable systems and the effect of a partial measurement is
done in reference \cite{Ata14c}.} or not. In the case of equation
\eqref{eq:psi_out_zou_wang_mandel_non-entangled} the answer is
affirmative, yielding the factorized form
\begin{equation}
\vert\psi_{out}\rangle=\left(-\sin\left(\frac{\varphi}{2}\right)\vert1_s\rangle
+\cos\left(\frac{\varphi}{2}\right)\vert1_{s'}\rangle\right)\otimes\vert1_i\rangle
\end{equation}
hence the interference pattern obtained in the singles detection
rate $P_s$ in equation \eqref{eq:P_s_singles_interference}. This
factorization is impossible in the case of equation
\eqref{eq:psi_out_ou_wang_zou_mandel_entangled}, therefore in this
case the subsystem of the signal beams only cannot be treated
separately, as proven by the missing interference pattern in
equation \eqref{eq:Ps_singles_rate_no_interf}.

\section{Imaging an object with undetected photons}
\label{sec:zeilinger_experiment}

In a recent experiment \cite{Lem14}, the same principle used by
Zou, Wang and Mandel was implemented in order to image an object
with undetected photons. The experiment (Figure 1 in \cite{Lem14})
is almost identical in principle to the one depicted in
Fig.~\ref{fig:zou_wang_mandel_experiment} and the object to be
imaged is placed in the dashed region on the beam $i_1$.
Therefore, depending on how the beam $i_1$ is attenuated and/or
dephased before entering the crystal $NL_2$, the
indistinguishability of the two idler beans will me more or less
perfect.

\begin{figure}
\centering
\includegraphics[width=2.8in]{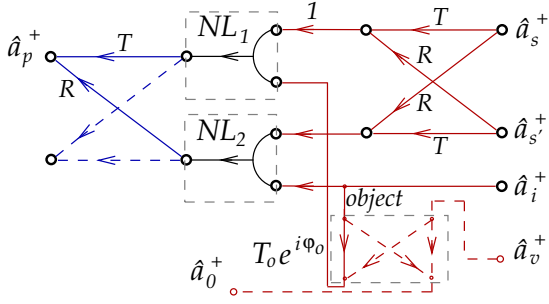}
\caption{(Color online) The graphical description of the Lemos
\emph{et al.} experiment \cite{Lem14}. Attenuation is QO is
modelled through a beam splitter. The two idler beams are
connected through the attenuation loss $T_o$ and phase shift
$\text{e}^{i\varphi_o}$ introduced by the object to be imaged.
However, the empty input mode $\hat{a}_0^\dagger$ as well as the
discarded output mode $\hat{a}_v^\dagger$ have to be taken into
account.} \label{fig:graphical_method_Lemos_Zeilinger_color}
\end{figure}

In Fig.~\ref{fig:graphical_method_Lemos_Zeilinger_color} we give
the graphical representation of this experiment. The idler beams
are now connected through a factor $T_o\text{e}^{i\varphi}$ where
$T_o$ and, respectively, $\text{e}^{i\varphi_o}$,  model the
transmittance and, respectively, the phase shift introduced by the
object intended to be imaged. For a perfectly transparent/opacque
object we have $T_o=1/0$. In order to have a unitary evolution,
the empty input mode $\hat{a}_0^\dagger$ as well as the
``absorbed'' output mode $\hat{a}_v^\dagger$ have to be taken into
account.
%
Inspecting the graph from
Fig.~\ref{fig:graphical_method_Lemos_Zeilinger_color} allows one
to write down the field operator transformation yielding
\begin{eqnarray}
\hat{a}_p^\dagger=T\left(T\hat{a}_s^\dagger+R\hat{a}_{s'}^\dagger\right)
\text{e}^{i\varphi_o}\left(T_o\hat{a}_i^\dagger+R_o\hat{a}_v^\dagger\right)
\nonumber\\
+R\left(R\hat{a}_s^\dagger+T\hat{a}_{s'}^\dagger\right)\hat{a}_i^\dagger
\end{eqnarray}
and assuming again balanced beam splitters we obtain the output
state vector
\begin{eqnarray}
\vert\psi_{out}\rangle=\frac{T_o\text{e}^{i\varphi_o}-1}{2}\vert1_s1_i\rangle
+\frac{R_o\text{e}^{i\varphi_o}}{2}\vert1_s1_v\rangle
\nonumber\\
+i\frac{T_o\text{e}^{i\varphi_o}+1}{2}\vert1_{s'}1_i\rangle
+i\frac{R_o\text{e}^{i\varphi_o}}{2}\vert1_{s'}1_v\rangle
\end{eqnarray}
Writing the output density matrix
$\hat{\rho}_{out}=\vert\psi_{out}\rangle\langle\psi_{out}\vert$
and tracing out the unused output ports ($s'$, $i$ and $v$) gets
us to
$\hat{\rho}_{s}=\text{Tr}_{s',i,v}\left\{\hat{\rho}_{out}\right\}$.
The probability of singles detection at $D_s$ is given by
\begin{eqnarray}
P_{s}=\text{Tr}\left\{\hat{a}_{s}^\dagger\hat{a}_{s}\hat{\rho}_{s}\right\}
=\frac{1-T_o\cos\left(\varphi_o\right)}{2}
\end{eqnarray}
where we used the fact that
$\vert{T_o}\vert^2+\vert{R_o}\vert^2=1$ and assumed $T_o$ real.
Therefore, both the transmissivity and the phase shift introduced
by the object modify the singles detection rate at $D_s$. For
example, for a transparent object we get
\begin{eqnarray}
P_{s}=\frac{1}{2}\left(1-\cos\left(\varphi_o\right)\right)
\end{eqnarray}
and a phase shift of $\pi$ radians can be clearly imaged, while a
$2\pi$ phase shift cannot. This is consistent with what was
experimentally found (Fig.~5 in \cite{Lem14}).

\section{Conclusions}
\label{sec:conclusions} In this paper we employed a graphical
method and a simple model of the spontaneous parametric
down-conversion in order to describe three experiments in quantum
optics.

The field operator transformations have been derived after a
simple inspection of the graphs describing the experiments. The
output state vectors have been obtained in a very fast and
straightforward manner. Despite the simple model used, we obtained
all the important features of these experiments, as described in
the respective papers.

\appendix
\section{Density matrix approach}
\label{sec:appendix_rho_operator} Having the output state vector
\eqref{eq:psi_out_ou_wang_zou_mandel_entangled} allows one to
compute the output density matrix operator
\begin{eqnarray}
\hat{\rho}_{out}=\vert\psi_{out}\rangle\langle\psi_{out}\vert
\end{eqnarray}
Since in this experiment we only use two output ports ($s$ and
$i$), we partially trace $\hat{\rho}_{out}$  over the two unused
outputs ($s'$ and $i'$) yielding
\begin{eqnarray}
\hat{\rho}_{s,i}=\text{Tr}_{s',i'}\left\{\hat{\rho}_{out}\right\}
=\sum_{m,n=0}^{\infty}{\langle{m}_{s'}n_{i'}\vert\psi_{out}\rangle\langle\psi_{out}\vert{m}_{s'}n_{i'}\rangle}
\end{eqnarray}
and after some straightforward calculations we arrive at the final
expression
\begin{eqnarray}
\label{eq:partial_rho_s_i_Ou_et_al}
\hat{\rho}_{s,i}=\frac{1}{2}\Big(\sin^2\left(\varphi'\right)\vert1_s1_i\rangle\langle1_s1_i\vert
+\cos^2\left(\varphi'\right)\vert1_{s}0_i\rangle\langle1_{s}0_i\vert
 \nonumber\\
+\cos^2\left(\varphi'\right)\vert0_{s}1_{i}\rangle\langle0_{s}1_{i}\vert
+\sin^2\left(\varphi'\right)\vert0\rangle\langle0\vert\Big)\quad
\end{eqnarray}

%
%

\end{document}